\documentclass[]{aa}
\usepackage{graphicx}
\usepackage{txfonts}
\begin{document}
\title{The emission line galaxy TV Reticuli
\thanks{Based on observations collected at the European Southern Observatory, 
        La Silla, Chile}}
\subtitle{Evidence for an ultraluminous supernova}
\titlerunning{The emission line galaxy TV Ret}
\author{L. Schmidtobreick \inst{1}
        \and
        C. Tappert \inst{2}
        \and
        H. Horst \inst{1,3}
        \and
        I. Saviane \inst{1}
        \and
        C. Lidman \inst{1}}
\offprints{Linda Schmidtobreick, \email{lschmidt@eso.org}}
\institute{European Southern Observatory, Casilla 19001, Santiago 19, Chile.
           \and
           Departamento de Astronom\'{\i}a y Astrof\'{\i}sica, 
           Pontificia Universidad Cat\'olica, Casilla 306, Santiago 22, Chile
           \and
           Zentrum f\"ur Astronomie, ITA, Universit\"at Heidelberg, 
           Albert-\"Uberle-Str. 2, 69120 Heidelberg, Germany}
\date{Received xxx xxx, xxx; accepted xxx xxx, xxx}
\abstract
{}
{TV\,Ret was classified as a cataclysmic variable due to an outburst
observed in 1977. We intended to confirm this classification 
and derive some basic properties of the system.}
{Low resolution optical spectra were obtained for a 
spectral classification of the object.}
{We find that the object is not a cataclysmic variable but an emission line
galaxy with a redshift $z=0.0964$. An R--image taken in very good 
seeing conditions shows that the object is extended.}
{We show that TV\,Ret is a blue dwarf galaxy, probably compact, with 
an absolute magnitude of  $M_B = -17.5$, a metallicity of 0.12 solar,
 and an average temperature of 
$1.3 \cdot 10^4$\,K. The line 
ratios place it among the H\,II galaxies, although close to the border of 
the Seyfert\,2s. The outburst, which was observed in 1977, could thus be 
explained by a supernova explosion. However, with an absolute magnitude around
$M_B = -21$, it was an extremely bright one.
}
\maketitle

\section{Introduction}
TV\,Ret was reported by  Kinman et al. (\cite{kinm+91}), who were
observing lightcurves of RR\,Lyr stars in the outer halo of the 
Large Magellanic Cloud. In one of the objects, which was named R1 in their
list, they noticed an increase in brightness, which they interpreted
as a possible dwarf nova outburst, and thus classified the object as a
cataclysmic variable.
As such, it was added in the 72nd name-list of variable stars as TV\,Ret
(Kazarovets \& Samus \cite{kaza+95}), and entered the catalogue of cataclysmic 
variables (Downes et al. \cite{down+01}).

We performed spectroscopic observations in order to confirm this 
classification. Surprisingly, the 
object turned out to be a narrow emission--line galaxy of nearly 0.1 redshift.
This class of galaxies consists of two different types:
starburst or H\,II-galaxies and the narrow--emission line AGNs, which 
are either Seyfert 2 galaxies or LINERs. 

In the first case, the emission lines origin in gas that is photo-ionised
by young, hot OB stars present in the star--forming regions. 
In the case of AGNs the emission lines are formed in the Narrow Line Region 
which is comprised of clouds ionised by radiation from the central engine
of the galaxy.
Several empirical methods have been developed
to distinguish between these two ionisation mechanisms, mainly by
comparing the line--ratios in the spectrum 
(see e.g. Baldwin et al. \cite{bald+81}; 
Veilleux \& Osterbrock, \cite{veil+87}; 
Dessauges-Zavadsky et al., \cite{dess+00}).

We attempt such a distinction and also derive the physical parameters
of the galaxy such as size, luminosity, metallicity, and gas 
temperature and density.

\section{Observations and data reduction}
\begin{table}[b]
\caption{\label{obstab} All observations obtained for this investigation 
listed with
their date, the telescope/instrument combination, the used grism and 
slit width, and the exposure time.}
\begin{tabular}{c c c c}
\hline
\hline
 \noalign{\smallskip}
Date & Telescope/Instrument & Grism/Slit & Exp-time [s] \\
 \noalign{\smallskip}
\hline
 \noalign{\smallskip}
2003-09-26 & 3.6m / EFOSC2 & Gr\#4 / 1.0$^{\prime\prime}$ & 1800 \\
2004-11-19 & 3.6m / EFOSC2 & Gr\#11 / 1.0$^{\prime\prime}$ & $3\times 900$ \\
 \noalign{\smallskip}
\hline
\end{tabular}
\end{table}

The observations were performed on 2003-09-26 and on 2004-11-19 using 
EFOSC2 (Buzzoni et al. \cite{buzz+84})
at the 3.6 m telescope at La Silla, Chile. 
In Table \ref{obstab} the observational
parameters are summarised. 

The standard reduction of the data was performed using IRAF. 
The bias was subtracted and the data were divided by a flat field,
which was normalised by fitting Chebyshev functions of high order. 
The spectra were optimally extracted (Horne \cite{horn86}).
Wavelength calibration yielded a final resolution of 1.1\,nm\ FWHM
for the 2003 data and 1.2\,nm\ FWHM for the 2004 data.
Flux calibration was performed only for the 2003 spectrum, using the
spectrophotometric standard LTT\,7379 which was observed  
with an airmass difference of 0.05. 

For all further analysis, if not especially indicated,
the MIDAS package and self--written routines were used.

\section{Results}
\begin{figure}
\rotatebox{-90}{\resizebox{!}{8.7cm}{\includegraphics{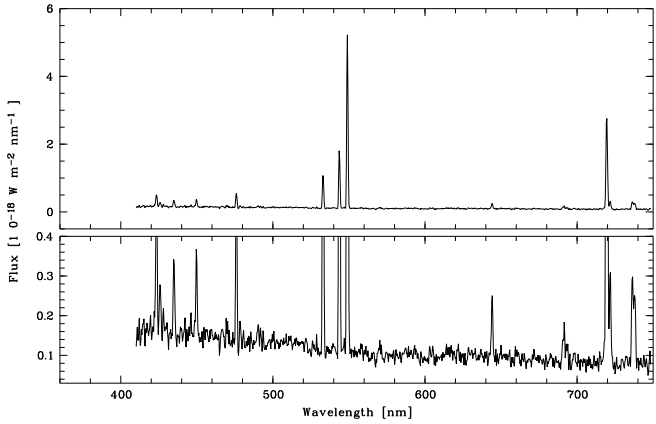}}}
\caption{\label{spec1} The flux--calibrated spectrum of TV Ret taken 
in September 2003. 
The upper plot shows the full intensity range;
in the lower plot the intensity range has been decreased to
better display the weak emission lines and the slope of the continuum.}
\end{figure}

\begin{figure}
\rotatebox{-90}{\resizebox{!}{8.7cm}{\includegraphics{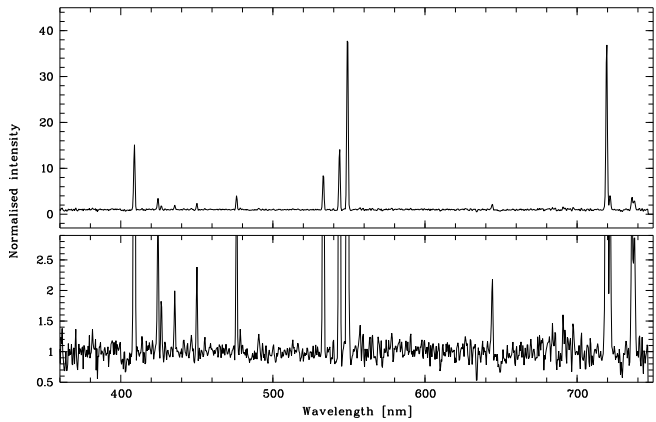}}}
\caption{\label{spec2} The spectrum of TV Ret taken in November 2004 
with the continuum
normalised to unity. The upper plot shows the full intensity range; 
in the lower plot the intensity range has been decreased to 
better display the weak emission lines.}
\end{figure}

\begin{figure}
\resizebox{8.7cm}{!}{\includegraphics{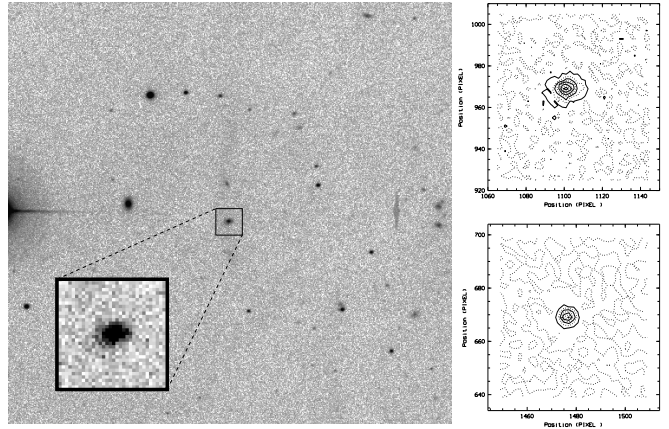}}
\caption{\label{finding} The R--band acquisition image shows the 
$3 \times 3$\,arcmin 
surroundings of TV\,Ret (in the centre) and a zoom on the object; 
north is up, east is left. 
With the good seeing of 0.6" the object appears clearly extended.
On the right side, the contours of TV Ret (upper plot) and a star
for comparison are plotted.}
\end{figure}

\begin{table*}
\caption{\label{linetab} Observed wavelengths, equivalent widths, 
and FWHM of all identified emission
lines in the 2002 and 2003 spectrum of TV\,Ret. For 2003 the line 
flux is also given. 
Note that the uncertainty of the line flux describes the 
uncertainty of the relative flux in the line and does not include the
photometric error. The line flux has been dereddened using the reddening laws
for H\,II regions and for AGNs (see text for details).}
\begin{tabular}{l r r r r r r r r r}
\hline
\hline
 \noalign{\smallskip}
           & \multicolumn{6}{c}{September 2003} & ~ ~ ~ ~ & \multicolumn{2}{c}{November 2004}\\
Transition & 
\multicolumn{1}{c}{$\lambda_{\rm obs}$ [nm]} & 
\multicolumn{1}{c}{$-W$ [nm]}  &
\multicolumn{1}{c}{$ F [10^{-18}$\,W\,m$^{-2}]$} & 
\multicolumn{1}{c}{$F_{\rm SB\,dered}$} & 
\multicolumn{1}{c}{$F_{\rm AGN\,dered}$} & 
\multicolumn{1}{c}{$ 100\cdot F_{\rm SB\,dered} / F_{\rm H\beta}$} & &
\multicolumn{1}{c}{$\lambda_{\rm obs}$ [nm]} & 
\multicolumn{1}{c}{$-W$ [nm]} \\
 \noalign{\smallskip}
\hline
 \noalign{\smallskip}
 H$_\alpha$   & 719.47 & 46.0 & 3.51(3) & 4.55 & 3.85 & 285(10) & & 719.42 & 46.1 \\
 H$_\beta$    & 532.97 & 10.5 & 1.08(2) & 1.60 & 1.24 & 100(4) & & 533.16 & 9.0 \\
 H$_\gamma$   & 475.97 &  3.5 & 0.46(2) & 0.72 & 0.54 &  45(3) & & 476.14 & 3.4 \\
 H$_\delta$   & 449.76 &  1.8 & 0.26(5) & 0.42 & 0.31 &  26(6) & & 450.01 & 1.4 \\
 H$_\epsilon$ & 434.94 &  1.8 & 0.26(2) & 0.42 & 0.31 &  26(3) & & 435.48 & 1.2 \\
 H$_8$        & 425.83 &      & & & & & & 426.68 \\
{[}S\,II]: 673.1  & 737.81 &  3.1 & 0.24(2) &      &      &  15(2) & & 737.58 &     \\
{[}S\,II]: 671.7  & 736.36 &  3.7 & 0.29(2) & 0.68 & 0.58 &  18(2) & & 736.20 & 6.8 \\
{[}N\,II]: 658.4  & 721.70 &  3.6 & 0.28(3) & 0.36 & 0.31 &  23(3) & & 721.64 & 3.9 \\
{[}O\,I]: $630.0^{(1)}$&691.20&2.3& 0.17(2) & 0.22 & 0.19 &  14(2) & &        & 1.6 \\
He\,I: 587.56 & 644.12 &  2.2 & 0.20(2) & 0.27 & 0.22 &  17(2) & & 644.20 & 2.0 \\
{[}O\,III]: 500.7 & 548.95 & 47.2 & 5.23(2) & 7.61 & 5.97 & 477(15) & & 549.04 & 50.4 \\
{[}O\,III]: 495.9 & 543.69 & 16.0 & 1.73(2) & 2.53 & 1.98 & 159(6) & & 543.82 & 17.1 \\
{[}O\,III]: 436.3 & 478.53 &  0.6 & 0.07(1) & 0.11 & 0.08 &   7(1) & & 478.57 & 0.5 \\
{[}Ne\,III]: 386.8& 423.44 &  3.1 & 0.48(3) & 0.78 & 0.57 &  49(4) & & 424.42 & 3.0 \\
{[}O\,II]: 372.7  &        &      &         &      &      & 169(15) & & 408.93 & 16.2\\

 \noalign{\smallskip}
\hline
\end{tabular}\\
(1): blend of several O\,I lines.

\end{table*}

\subsection{Classification}
In Fig.~\ref{spec1} the flux--calibrated spectrum observed in 2003
is plotted. The spectrum is clearly dominated by strong emission lines.
The continuum matches the spectral energy distribution (SED) of a late A or 
early F type 
star; the corresponding temperature is $T_{\rm eff} = 7500$\,K. 
In Fig. \ref{spec2}, 
the normalised spectrum from 2004 is plotted, which shows the same 
emission lines as the 2003 spectrum and an additional line at 408.9\,nm,
which was outside the spectral range of the 2003 data.
Both spectra show clearly that the object is
not a cataclysmic variable, as the typical emission lines for this kind
of object are not present at their rest wavelengths. 
Instead, we find several strong emission lines, 
which turned out to be redshifted Balmer lines as well as 
O\,I, O\,III, N\,II, S\,II, and Ne\,III. The properties of these lines
are given in Table~\ref{linetab}. 
We averaged the individual shifts of these
lines to find the redshift of the object as $z = 0.0964(2)$.

The redshift, which indicates an extragalactic object, as well as the 
strength of the ionised lines, are best interpreted if we assume the 
object to be an emission line galaxy,
which, however, is unresolved in previous images.
In Fig.~\ref{finding}, a $3\times 3$\,arcmin R--image is plotted, 
obtained under good seeing conditions of 0.6\,arcsec. 
In this image, the 
object appears extended and slightly elongated. The size can be estimated to 
about 0.7$\times$0.9\,arcsec. 

\subsection{Physical properties of the galaxy}

For all further calculations, we use the standard flat cosmology with
$\Omega_M = 0.27$ and $H_0 = 72\,\rm km\,s^{-1}\,Mpc^{-1}$
(Spergel et al. \cite{sper+03}). 

From $z = 0.0964(2)$ 
we derive the angular size distance 
$D_{\rm ang} = 358$\,Mpc. Thus, assuming that we see most of it, 
the size of the galaxy is 
$1.2\times 1.6$\,kpc, which matches the size of a rather small
dwarf galaxy.

The magnitude of TV\,Ret is given as $B=20.5$ (Kinman et al. \cite{kinm+91}). 
The absolute B--magnitude can be derived using
\begin{equation}
M_B = B - 5\log{D_{\rm lum}} +5 -K
\end{equation}
with the luminosity distance calculated as $D_{\rm lum}=430$\,Mpc,
and the k--correction 
$K = -2.5 \log{\left((1+z)^{-1} L_{\lambda (1+z)^{-1}} / L_{\lambda}\right)}$
(e.g. Oke \& Sandage \cite{oke+68}).
We use the slope 
$L_{\rm 401nm}/L_{\rm 440nm} = 1.15$ as derived from the spectrum 
plotted in Fig.\ref{spec1},
and derive the absolute magnitude
of the galaxy as $M_B = -17.5$. This value is typical of massive
dwarf galaxies (see e.g. Melisse \& Israel \cite{meli+94}) and thus confirms
the above classification. However, the 
surface brightness of 
$\sigma_B = 20.0$\,mag/arcsec$^2$
is rather high. 


\subsubsection{Starburst or AGN?}
\begin{table}
\caption{\label{loglinerat} The logarithm of the line ratios is given for 
the two reddening laws, starburst (SB) and AGN, 
as derived from $\rm H\alpha / H\beta$.
The forbidden line fluxes correspond to the following transitions: 
O\,III: 500.7\,nm, N\,II: 658.3\,nm, S\,II: 671.6\,+\,673.1\,nm, and
O\,I: 630.0\,nm.}
\begin{tabular}{c c c c c c}
\hline
\hline
 \noalign{\smallskip}
 & $\rm \log{\frac{[O\,III]}{H\beta}}$ & $\rm \log{\frac{[N\,II]}{H\alpha}}$ & $\rm \log{\frac{[S\,II]}{H\alpha}}$ & $\rm \log{\frac{[O\,I]}{H\alpha}}$ & $\rm \frac{H\alpha}{H\beta}$ \\
 \noalign{\smallskip}
\hline
 \noalign{\smallskip}
SB  & 0.68 & -1.10 & -0.83 & -1.32 & 2.85 \\
AGN & 0.68 & -1.09 & -0.82 & -1.31 & 3.10 \\
 \noalign{\smallskip}
\hline
\end{tabular}
\end{table}

A more detailed classification of the galaxy can be done by analysing the 
emission lines.
Apart from the features at 478.5\,nm and 737\,nm, which are clearly blends of 
{[}O\,I] and {[}SII\,673.1]/{[}SII\,671.7] respectively, the lines are not resolved 
within our resolution of 1.1\,nm, 
and therefore are narrow--emission lines. 
Thus, TV\,Ret could be either an H\,II galaxy or a narrow--line AGN.
We use the flux--ratio of the different emission lines to find the
mechanism by which the emission lines are produced following the method 
described by Veilleux \& Osterbrock (\cite{veil+87}).

To have a comparable data set, we also determined the reddening in the 
same way as these authors, using the 
H$\alpha$/H$\beta$ flux--ratio and the Whitford reddening curve parameterised 
by Miller \& Mathews (\cite{mill+72}). We measure 
$F({\rm H\alpha}) / F({\rm H\beta}) = 3.25$. If we assume
a starburst galaxy with a recombination value 
$I({\rm H\beta}) / I({\rm H\alpha})=2.85$, the reddening 
is $E(B-V) = 0.13$. If we assume an AGN with 
$I({\rm H\beta}) / I({\rm H\alpha})=3.1$, we find a rather low value of
$E(B-V) = 0.05$. For both cases, the dereddened flux values are listed
in Table~\ref{linetab}. Although the dereddened flux values are different,
the logarithms of the line ratios used for the classification are similar 
(see Table~\ref{loglinerat}). This is expected, as the line ratios were chosen 
for being insensitive to the reddening, i.e. are close in wavelength.

We compared the line ratios of the forbidden lines (see Table~\ref{loglinerat})
with the values given by Veilleux \& Osterbrock 
(\cite{veil+87}), i.e. their figures 1--6. In all these diagnostic diagrams,
TV\,Ret lies close to the
border between AGNs and H\,II region--like objects, which is mainly due 
to the high value of 
$\rm {[}O\,III]\lambda 500.7 / H\beta$.
In $\rm {[}O\,III] / H\beta$ versus 
$\rm {[}N\,II] / H\alpha$ and $\rm {[}O\,III] / H\beta$ 
versus $\rm {[}S\,II] / H\alpha$, 
it lies on the side of the H\,II region--like objects, while for
$\rm {[}O\,III] / H\beta$ versus $\rm {[}O\,I] / H\alpha$
it lies on the side of the AGNs. 

Comparing the line ratios with the diagnostic diagrams of 
Dessauges-Zavadsky (\cite{dess+00}) yields similar conclusions.
We can definitely exclude that TV\,Ret is a LINER, but it is on the 
border between the Seyfert 2 and the starburst galaxies, although the 
latter appears slightly more likely.

\subsubsection{Temperature and metallicity}

\begin{table}
\caption{\label{parameter} Physical parameters of the HII galaxy}
\begin{centering}\begin{tabular}{rrl}
&
&
\tabularnewline
\hline
\hline
\multicolumn{1}{c}{Quantity }&
\multicolumn{1}{c}{Value}&
\multicolumn{1}{c}{Notes}\tabularnewline
\hline
$T({\rm OIII})$&
$13400_{938}^{969}$ K &
\tabularnewline
$n({\rm SII})$&
$239_{177}^{433}$ cm$^{-3}$ &
\tabularnewline
${\rm O}^{++}/{\rm H}^{+}$ &
$(7.02\pm0.19)\cdot 10^{-5}$ &
average $\lambda5007$ and $\lambda4959$ \tabularnewline
${\rm O}^{+}/{\rm H}^{+}$&
$(2.39\pm0.90)\cdot 10^{-5}$ &
$\lambda3727$\tabularnewline
${\rm O/H}$  &
$(9.41\pm0.92)\cdot 10^{-5}$ &
TOT\tabularnewline
$12+\log({\rm O/H})$ &
$7.97\pm0.04$ &
\tabularnewline
${\rm N+/O+}$&
$(9.87\pm6.48)\cdot 10^{-2}$ &
\tabularnewline
$\log({\rm N/O})$ &
$-1.01\pm0.22$&
\tabularnewline
${\rm N/H}$&
$(2.36\pm2.44)\cdot 10^{-6}$&
\tabularnewline
${\rm S}^{+}/{\rm H}^{+}$ &
$(5.69\pm0.28)\cdot 10^{-7}$&
average $\lambda6716$ and $\lambda6731$\tabularnewline
\hline
\end{tabular}\par\end{centering}
\end{table}

For a first indication, we compared the line ratios with the models of 
Ferland \& Netzer (\cite{ferl+83}) and Evans \& Dopita (\cite{evan+85}),
which are overplotted on the diagnostic diagrams of Veilleux \& Osterbrock
(\cite{veil+87}), and find a metallicity 
of 0.1 times solar and an ionisation temperature of 45000\,K. 

A more sophisticated computation of the abundances have been done 
following the so-called direct method
(e.g. Osterbrock \cite{oste89}). The {[}OIII] region electron
temperature $T_{{\rm e}}$, electron density $n_{{\rm e}}$, and abundances 
were computed using tasks within the NEBULAR package of IRAF.
The temperature was computed using the {[}OIII] lines ratio 
$(4959+5007)/4363$,
and the density using the {[}SII] line ratio $6716/6731$. Using these 
values for $T_{{\rm e}}$ and $n_{{\rm e}}$, the ionic abundances 
were computed with the IONIC task, with central wavelengths, 
line ratios and errors taken from Table~\ref{linetab}. 
The total oxygen abundance
was computed as ${\rm O/H=O^{++}/H^{+}+O^{+}/H^{+}}$, i.e. neglecting 
the usually small contribution from ${\rm O^{+3}}$ (Skillman \& Kennicutt
\cite{skil+93}). In the case of sulfur, the abundance
was computed as ${\rm S/H = ICF \times (S^{+}+S^{++})}$, 
with the ionisation correction factor (ICF) computed as 
${\rm ICF=[1-(1-O^{+}/O)^{\alpha}]^{-1/\alpha}}$.
This expression for the ICF was first proposed by Stasinska
(\cite{stas78})
with $\alpha=3$, but we used the value $\alpha=2.6$ which reproduces
Garnett's (\cite{garn89}) photo-ionisation models better in the
range ${\rm O^{+}/O>0.2}$. Nitrogen abundances were computed
assuming ${\rm (N/O)=(N^{+}/O^{+})}$, and then using the nitrogen
to oxygen ratio in the equation ${\rm (N/H)=(N/O)}\times{\rm (O/H)}$.

In Table~\ref{parameter}, the derived parameters are listed for TV\,Ret.
We find that the average metallicity is about 0.12 solar, in agreement 
with the values derived from the model of Ferland \& Netzer (\cite{ferl+83}).
The gas temperature derived from the {[}OIII] lines is $1.3(1)\cdot 10^4$\,K, 
the density derived from the
{[}S\,II] lines is about 240/cm$^3$.
These values again confirm TV\,Ret as an H\,II galaxy.

\subsection{The outburst}
\begin{figure}
\rotatebox{-90}{\resizebox{!}{8.7cm}{\includegraphics{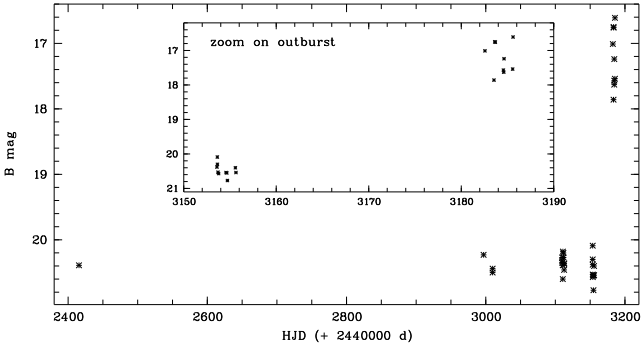}}}
\caption{\label{lc} The B--magnitude of TV\,Ret plotted against
the heliocentric Julian Date. The data are taken from Table 5 of
Kinman et al. (\cite{kinm+91}). The smaller plot is a zoom on the 
last 40 days, when the outburst happened.
}
\end{figure}

The reason for the classification of TV\,Ret as a cataclysmic variable
was the outburst that the object underwent in February 1977 and that
was observed over several days (Kinman et al. \cite{kinm+91}). 
In Fig.~\ref{lc}, the B--magnitudes and heliocentric Julian Date
that they give in Table 5 are plotted. They estimated the
outburst amplitude as $\Delta m = 3.8$\,mag. 
Note that there is no tight 
observational constraint on the duration of the outburst. The rise happened
between JD\,2443156 and JD\,2443182, and hence lasted 26 days at the most. 
However, the decline was not observed, and thus the actual duration of the
outburst is unknown. We can only
say that the object is back in quiescence during our observations 26 years 
later.

If this was an outburst in a foreground object, by chance superimposed on the
galaxy,
we can estimate an upper limit for the brightness of this object. 
With a $S/N\approx10$ of our spectra, we would be able to see an 
object about three times
fainter than the galaxy. Since no trace of such a forground object is 
visible in the spectrum, it must be fainter than $\approx$22\,mag.
In that case, the 
amplitude of the outburst was at least 5.5\,mag, leaving two 
explanations, a dwarf nova or a nova outburst. In the case of a dwarf nova, the
object would be in quiescence now, close to 22\,mag, but would show strong 
Balmer emission lines.
No emission is found at the restframe wavelength of these lines, making
this possibility very unlikely. Note that  we cannot refute the possibility
of a dwarf--nova super--outburst which would imply that the quiescent object 
is even fainter, and the Balmer emission are no longer observable.
In the case of a nova, the outburst amplitude itself would have been larger, 
so that the object is not visible in quiescence. However, even faint novae 
have an absolute magnitude of $M_B = -7$ (e.g. Della Valle \& Livio, 
\cite{dell+95}), yielding a lower limit of its
distance as 450\,kpc and placing it far outside our Galaxy.

In the following, we assume that the outburst originates in the galaxy
and is not a chance transient from a foreground object.  The
luminosity $L_{\rm out}$ of the outburst can then be estimated via
\begin{equation}
\Delta m = -2.5 \log{\frac{L_{\rm gal} + L_{\rm out}}{L_{\rm gal}}}
\end{equation}
and thus results in $L_{\rm out} = 32.1\,L_{\rm gal}$.
The absolute B--magnitude at the maximum is $M_B = -21.3$.  While the
explanation of a supernova seems natural, the luminosity of the
outburst is too high. The brightest supernovae of Type Ia
have an absolute B--magnitude around $M_B = -19.8$ (e.g.  Germany et
al.~\cite{germ+04}). Even if we allow for a large K--correction of
0.4\,mag (e.g. Hamuy et al. \cite{hamu+93}), the outburst would still
be about 1\,mag too bright for a Type Ia supernova. Furthermore,
supernovae are not known to vary on short time scales. The lightcurve
observed by Kinman et al., instead, varies around 0.8\,mag and on time
scales of hours both during quiescence and during outburst (see
Fig.~\ref{lc}).  
There are however unusual supernovae that are
extremely luminous. The brightest one is SN 1999as (Knop et
al. \cite{knop+99}), a type-Ic hypernova that reached an absolute V
magnitude brighter than -21, i.e. similar to the magnitude that was
found for TV\,Ret.  However, the probability for such objects is
rather low.  
Even more extreme are pair-production supernovae
(Scannapieco et al., \cite{scan+05}), which can reach $M_B = -21$. However,
pair-production supernovae have never been observed and they are
thought to occur in environments with metallicities that are several
orders of magnitude lower than the metallicity of the host galaxy of
TV\,Ret.

The variation on short time scales indicates a compact source, so an AGN
seems more likely. However, they do not generally show such strong variations
in the optical.
While X-ray variability on short timescales is a common phenomenon in AGNs,
optical intra-night variability of up to 10\% is only found in luminous 
Quasars with $M_B < -24.5$ 
(Gupta \& Joshi \cite{gupt+05}; Stalin et al. \cite{stal+05}). 
More extreme variabilities are only exhibited in Radio-loud AGN with 
ultra--relativistic jets and there is no indication that TV\,Ret is 
such an object. Radio and/or x-ray observations of this galaxy 
would help to decide
on the presence of an AGN and thus also on the possible nature of the
outburst.

There is a possible third explanation, if we assume that the accuracy of the
measurements of Kinman et al. is not the 0.03\,mag that they claim, but
rather of the order of 0.3\,mag. In that case, the short--term variation 
would not be real
and for the outburst magnitude one would have to subtract the average 
magnitude values of outburst and quiescence magnitudes instead of the 
extreme values. 
This would lower the
outburst amplitude by 0.4\,mag, putting it closer to the possible range of
a bright supernova. However, the method that the authors used is 
well--known and the quoted errors seem reasonable for it. 
Still, the fact that the variation is also
present during the quiescence phase might indicate that it is a scatter
in the measurements rather than a real variation. On the other hand, the 
scatter seems to be slightly larger during outburst, where one would actually
expect the higher accuracy. 
We know of no photometric monitoring of TV\,Ret after 1977, which would
be an important observation. If the
short--term variation was confirmed by such measurements, this would be
a strong indication for the presence of a compact source, i.e. an AGN. 
If no variation was detected, the probability would grow that the 
variation in 1977 is just the uncertainty in the measurements.

All in all, the reason for the outburst remains a mystery. Too many possible
explanations exist, and none of them is really convincing. Unless the
original measurements have rather high uncertainties, we can 
conclude that the outburst was some rare and unusual event that is not 
comparable with normal sources of variability. 

\section{Conclusions}
We refute the classification of TV\,Ret as a cataclysmic variable.
The object is a narrow--emission line galaxy, probably of H\,II type, although
we cannot rule out the contribution of an AGN to the ionisation field.
The size of $\approx$1.2\,kpc and the absolute magnituse of $M_B = -17.5$
places the object among the dwarf galaxies. 
Assuming that we see most of the galaxy, it is probably a blue compact dwarf 
with ongoing star formation. 
The metallicity is 0.12 solar, in agreement with an H\,II galaxy.

We have no final conclusion concerning the observed outburst. If associated 
with the galaxy, it is about 1\,mag brighter than normal supernovae of type Ia.
One would thus prefer to explain the outburst via a foreground transient.
However, 
the two possible explanations -- dwarf nova or nova -- have their drawbacks as
well.

\acknowledgement{ 
This research has 
made use of the Simbad database operated at CDS, Strasbourg, France.
We thank the referee for valuable comments.
}

\end{document}